\documentclass[twocolumn,amsmath,amssymb,showpacs]{revtex4}
\usepackage{graphicx}
\usepackage{longtable}
\usepackage{dcolumn}
\usepackage{bm}
\usepackage[plainpages = false, pdfpagelabels, 
                 bookmarks,
                 bookmarksopen = true,
                 bookmarksnumbered = true,
                 linktocpage,
                 pagebackref=false,
                 colorlinks = true,
                 linkcolor = blue,
                 urlcolor  = blue,
                 citecolor = blue,
                 anchorcolor = green,
                 hyperindex = true,
                 hyperfigures
                 ]{hyperref} 
\begin{document}

\title {Charge-changing cross section and interaction cross section for 4$\leq$Z$ \leq$9 isotopes}

\author{M. \surname{Imran}$^1$}
\author{Z. \surname{Hasan}$^2$}
\author{A. A. \surname{Usmani}$^1$}
\author{Z. A. \surname{Khan}$^1$}\email{zakhan.amu@gmail.com}
\affiliation{$^1$Department of Physics, Aligarh Muslim University, Aligarh-202002, India}
\affiliation{$^2$Department of Applied Physics, ZH College of Engineering and Technology, Aligarh Muslim University, Aligarh-202002, India}

\begin{abstract}
The root-mean-square proton and neutron radii for $^{7,9-12,14}$\rm Be, $^{10-15,17}$\rm B, $^{12-19}$\rm C, $^{14,15,17-22}$\rm N, $^{16,18-24}$\rm O, and $^{18-21,23-26}$\rm F isotopes are deduced from a systematic analysis of experimental charge-changing and interaction cross sections in the framework of Glauber model. The calculations involve descriptions of nuclei based on Slater determinants using harmonic oscillator single-particle wave functions. The extracted proton and neutron radii have been examined in the light of some important features such as neutron skin thickness/halo-like structure/subshell closure observed in exotic isotopes.
 
\end{abstract}
\maketitle

\section{Introduction}
\label{sec1}

The production and study of unstable nuclei away from the stability line has been a source of new impetus to the field of both nuclear
physics and nuclear astrophysics. The important aspect of such nuclei is the existence of thick neutron skins and halos in neutron-rich nuclei 
\cite{1,2,3}. These exotic features of unstable neutron rich nuclei may result from the presence of neutron dominated envelop in the nuclear surface 
region. Here, it may be mentioned that a reliable information about the proton radii of such nuclei is also a matter of concern before 
extracting the neutron skin thickness and understanding the nuclear halo which involves a large spatial separation of one or two weakly bound
valence neutrons, thereby forming a low-density neutron halo around the core nucleus. It is well known that electron scattering is an ideal tool 
for probing the proton (charge) distribution in stable nuclei, but this very approach has  been utilized so far for limited unstable (short lived) 
nuclei. Isotope shift measurements though help us to deduce the proton radii but they are also limited to only few unstable nuclei. Alternately, the 
measurement of the charge-changing cross sections (CCCS) \cite{4,5,6,7,8,9,10} may find its place to get information about the proton radii of unstable nuclei. 
One hopes that the combined study of charge-changing and the corresponding reaction (interaction) cross sections could be helpful in 
providing reliable estimates for the proton and matter radii of unstable nuclei.

It has been demonstrated that the Glauber model has been quite successful in extracting the matter radii of radioactive nuclei \cite{11} from the 
corresponding experimental data on the reaction cross sections, with some reasonable adjustments in proton and neutron radii. In the case of
CCCS, it was thought that these cross sections may involve only the proton density of the projectile, and hence it was
expected that the analysis of CCCS may directly provide the information about the proton radii of unstable nuclei. However, calculations of CCCS using the 
Glauber model \cite{5,12,13,14,15,16} suggest that it is not only the protons in the projectile, the presence of neutrons also matters in explaining the 
charge-changing cross section data. Due to this, it becomes difficult to understand the reaction mechanism for charge-changing cross sections.
In order to accomodate CCCS within the framework of the Glauber model, it has, however, been suggested that calculations of CCCS may be possible by introducing a phenomenological correction parameter
\cite{5,12,17} that takes care of the presence of neutrons in the projectile. To appreciate the motive of the present work, it is necessary to comment on
the correction parameter \cite{5,12,17} used in the study of CCCS. As noticed in Refs. \cite{5,17}, although the correction parameter was parametrized as a function
of the incident energy, it shows weak energy dependence in the energy range 200-600 MeV/nucleon,  giving rise to almost a constant value ($\sim$ 1.107) for each of the
isotopic chain. On the other hand, in Ref. \cite{12}, the correction parameter is related to the ratio Z/N (Z = atomic number, N = neutron number) of the projectile without reference to the energy, thereby 
using its different value for each of the isotopes of a given element. 
Apart from the energy factor, the difference between
the two approaches is the involvement of density distributions in the calculations of CCCS: the approach in Refs. \cite{5,17} involves only the proton 
density of the projectile, whereas the calculations in Ref. \cite{12} require both the proton and neutron densities. In our opinion, since the proton
radii are crucial for obtaining the neutron skin (surface) thickness and understanding the halos in neutron rich unstable nuclei, the approach of 
involving only the proton densities, along with a correction parameter, seems to be a better choice for exploiting the CCCS. Once the proton radius
is obtained for a given nucleus, the neutron radius is then obtained from the analysis of the corresponding reaction (interaction) cross section data. In
this way, one hopes to provide a better understanding of proton (charge) and matter radii of neutron rich unstable nuclei. 

In this work, we propose a different prescription to get the correction parameter, needed to incorporate the contribution due to the presence of projectile neutrons to CCCS.
Explicit calculations have been carried out for CCCS for beryllium, boron, carbon, nitrogen, oxygen, and fluorine isotopes within the framework of 
correlation expansion for the Glauber model S-matrix \cite{18} by considering up to the two-body correlation term. The basic input of the Glauber 
model, the nucleon-nucleon ($NN$) amplitude, considers the nuclear in-medium effects, and the nuclear densities are obtained using the Slater
determinants consisting of the harmonic oscillator single-particle wave functions (hereafter referred to as SDHO densities) which involve
oscillator constant as their input parameter. The proton densities obtained from the charge-changing cross section data are then used to reproduce the corresponding 
reaction (interaction) cross sections with the adjustment of the oscillator constant in SDHO  neutron density distributions. The extracted proton and neutron radii have been examined in view of some important features such as neutron skin thickness/halo-like structure/subshell closure observed in neutron rich unstable nuclei. 

Finally, we have performed parameter free calculations to test the reliability of the present method used to deduce the proton and neutron radii (densities) in other situations. For this we have analysed the CCCS for $^{12-19}$\rm C and $^{14,15,17-22}$\rm N isotopes on a proton target using the aforesaid information about the proton and neutron radii (densities), and the results are discussed in the light of some recent calculations \cite{19} .
                        
The formulation of the problem is given in  Sec. \ref{sec2}. The numerical results are presented and discussed in Sec. \ref{sec3}. The conclusions are 
presented in Sec. \ref{sec4}.

\section{Formulation}
\label{sec2}
\subsection{Nucleus-nucleus reaction (interaction) cross section}

According to the Glauber model, the reaction cross section ($\sigma_{R}$) for the scattering of a projectile nucleus with ground state 
wave function $\psi_{P}$ on a target nucleus with ground state wave function $\psi_{T}$ is given by
\begin{equation}
\sigma_{R}= 2\pi\int \left[1- \vert S_{el}(b)\vert^{2}\right]b~db,
\label{eq1}
\end{equation}
\begin{equation}
S_{el}(b)= \langle \psi_{T}\psi_{P}\vert\prod^{A}_{i=1}\prod^{B}_{j=1}[1-\Gamma_{NN}(\vec{b}-\vec{s_{i}}+\vec{s^{'}_{j}})]\vert\psi_{P}\psi_{T}\rangle,
\label{eq2}
\end{equation}
where $A(B)$ is the mass number of target (projectile) nucleus, $\vec{b}$ is the impact parameter vector perpendicular to the incident momentum 
, $\vec {s_{i}}\vec{(s^{'}_{j}})$ are the projections of the target (projectile) nucleon coordinates on the impact parameter plane, 
and $\Gamma_{NN}({b})$ is the $NN$ profile function, which is related to the $NN$ scattering amplitude $f_{NN}(q)$ as 
follows  
\begin{equation}
\Gamma_{NN}({b})= \frac{1}{2\pi ik}~\int \exp(-i\vec {q}.\vec {b})f_{NN}({q})~d^{2}q, 
\label{eq3}
\end{equation}
where $k$ is the incident nucleon momentum corresponding to the projectile kinetic energy per nucleon, and $\vec{q}$ is the momentum transfer.

Following Ahmad \cite{18}, the S-matrix element, $S_{el}({b})$, upto two-body correlation term takes the following form:
\begin{equation}
S_{el}({b})\approx S_{0}(b)+S_{2}(b),
\label {eq4}
\end{equation}
where
\begin{equation}
S_{0}({b})= [1-\Gamma^{NN}_{00}(b)]^{AB}, 
\label{eq5}
\end{equation}
and
\begin{gather}
S_{2}({b})=\Bigl\langle\psi_{T}\psi_{P}\Bigl|\frac{1}{2!}[1-\Gamma^{NN}_{00}(b)]^{AB-2}~~~~~~~~~~~~~~~~~~\nonumber\\
~~~~~~~~~~~~~~~~~~\times \sum^{'}_{i_{1},j_{1}}\sum^{'}_{i_{2},j_{2}}\gamma_{i_{1},j_{1}}\gamma_{i_{2},j_{2}}\Bigr|\psi_{P}\psi_{T}\Bigr\rangle,
\label{eq6}
\end{gather}
with 
\begin{equation}
\gamma_{ij}=\Gamma^{NN}_{00}(\vec{b})-\Gamma_{NN}(\vec{b}-\vec{s_{i}}+\vec{s^{'}_{j}}), 
\label{eq7}
\end{equation}
and
\begin{equation}
\Gamma^{NN}_{00}({b})=\int \rho_{T}(\vec r)~\rho_{P}(\vec{r^{'}})~\Gamma_{NN}(\vec b-\vec s +\vec{s^{'}})~d\vec r~d\vec{r^{'}}.
\label {eq8}
\end{equation}
The primes on the summation signs in Eq. (\ref{eq6}) indicate the restriction that two pairs of indices can not be equal at the same time (for example,
if $i_{1}$=$i_{2}$ then $j_{1} \neq j_{2}$ and vice versa). The quantities $\rho_{T}$ and $\rho_{P}$ in Eq. (\ref{eq8}) are the (one-body) ground state
densities of the target and projectile, respectively. Here, it is to be noted that the distinction between protons and neutrons in both the projectile and target has been included in the uncorrelated part ($S_{0}$) of the S-matrix element only. Such a distinction involves different values of the parameters for pp and pn scattering amplitudes, and uses different density distributions for protons and neutrons in the colliding nuclei. With this consideration, $S_{0}(b)$ takes the form
\begin{gather}
S_{0}({b})= [1-\Gamma^{pp}_{00}({b})]^{Z_{P}Z_{T}}[1-\Gamma^{np}_{00}({b})]^{N_{P}Z_{T}}~~~\nonumber\\ 
~~~~~~~~~~~~~~~~~\times[1-\Gamma^{pn}_{00}({b})]^{Z_{P}N_{T}}[1-\Gamma^{nn}_{00}({b})]^{N_{P}N_{T}},
\label{eq9}
\end{gather}
with
\begin{equation}
\Gamma^{ij}_{00}({b})=\int \rho^{j}_{T}(\vec r_{j})~\rho^{i}_{P}(\vec{r_{i}^{'}})~\Gamma_{ij}(\vec b-\vec s_{j} +\vec{s_{i}^{'}})~d\vec r_{j}~d\vec{r_{i}^{'}}
\label{eq10}
\end{equation}
where $Z_{T}(Z_{P})$ and $N_{T}(N_{P})$ are the target(projectile) atomic and neutron number, respectively, and each of $i$ and $j$ stands for a proton and a neutron.

Regarding the two-body correlation term ($S_{2}({b})$), it should be mentioned that the consideration of distinct features of protons and neutrons in this term is not as straightforward as it was in $S_{0}({b})$. Moreover, we may also add that (i) the two-body correlation term provides only a correction to the dominant part of the S-matrix element i.e. $S_{0}({b})$ and (ii) the contribution of this term is expected to be small. Due to this, we have used average values of the parameters for pp and pn amplitudes, and involve matter density distributions in the evaluation of $S_{2}({b})$. More explicitly, the evaluation of $S_{2}({b})$ gives the following form \cite{18}:
\begin{gather}
S_{2}({b})=-\frac{AB}{8\pi^{2}k^{2}}[1-\Gamma^{NN}_{00}(b)]^{AB-2}[(A-1)(B-1)\nonumber\\
~~~~~~~~~~~~~~~~~\times(G_{22}(b)-G_{00}^{2}(b))+(B-1)\nonumber\\
~~~~~~~~~~~~~~~~~\times(G_{21}(b)-G_{00}^{2}(b))+(A-1)\nonumber\\
~~~~~~\times(G_{12}(b)-G_{00}^{2}(b))],
\label{eq11}
\end{gather}
with
\begin{gather}
G_{22}(b)=\int \exp[-i(\vec q_{1}+\vec q_{2}).\vec b]F^{(2)}_{T}(\vec q_{1},\vec q_{2})F^{(2)}_{P}(  -\vec q_{1},-\vec q_{2})\nonumber\\
~~\times f_{NN}(q_{1})f_{NN}(q_{2})d^{2}q_{1}d^{2}q_{2},
\label{eq12}
\end{gather}
\begin{gather}
G_{21}(b)=\int \exp[-i(\vec q_{1}+\vec q_{2}).\vec b]F_{T}(\vec q_{1}+\vec q_{2})F^{(2)}_{P}( -\vec q_{1},-\vec q_{2})\nonumber\\
\times f_{NN}(q_{1})f_{NN}(q_{2})d^{2}q_{1}d^{2}q_{2},
\label{eq13}
\end{gather}
\begin{gather}	
G_{12}(b)=\int \exp[-i(\vec q_{1}+\vec q_{2}).\vec b]F^{(2)}_{T}(\vec q_{1},\vec q_{2})F_{P}( -\vec q_{1}-\vec q_{2})\nonumber\\
~~\times f_{NN}(q_{1})f_{NN}(q_{2})d^{2}q_{1}d^{2}q_{2},
\label{eq14}
\end{gather}
and
\begin{equation} 
\begin{aligned}
G_{00}({b}) & =\int \exp(-i\vec q.\vec b)F_{T}(q)F_{P}(q)f_{NN}(q)d^{2}q,\\ 
            & = (2\pi ik)\Gamma^{NN}_{00}(b).
\end{aligned}
\label{eq15}
\end{equation}
The quantities $F_{T(P)}(q)$ and $F^{(2)}_{T(P)}(\vec q_{1},\vec q_{2})$ in the above equations are the one- and two-body form factors of the target(projectile) nucleus, respectively,
\begin{equation}
F_{T(P)}(q)=\int \rho_{T(P)}(r) \exp(i\vec q.\vec r)d{\vec r},
\label{eq16}
\end{equation}
\begin{equation}
F^{(2)}_{T(P)}(\vec q_{1},\vec q_{2})=\int \rho^{(2)}_{T(P)}(\vec r_{1},\vec r_{2}) \exp(i\vec q_{1}.\vec r_{1}+\vec q_{2}.\vec r_{2})d{\vec r_{1}}d{\vec r_{2}},
\label{eq17}
\end{equation} 
where $\rho^{(2)}_{T(P)}(\vec r_{1},\vec r_{2})$ is the two-body density of the target(projectile) nucleus. It is clear that evaluation of $F_{T(P)}$, which requires (intrinsic) one-body density distribution of the target(projectile), is trivial. To evaluate $F^{(2)}_{T(P)}(\vec q_{1},\vec q_{2})$, we must know the intrinsic two-body density of the target(projectile) nucleus. For this, one starts with assuming a model wave function $\Phi_{C}$ and writes the intrinsic one- and two-body form factors as follows \cite{20}
\begin{equation}
F_{\nu}(q)=\theta_{\nu}(q)F^{C}_{\nu}(q),
\label{eq18}
\end{equation}
\begin{equation}
F^{(2)}_{\nu}(\vec q_{1},\vec q_{2})=\theta_{\nu}(\vec q_{1}+\vec q_{2})F^{C(2)}_{\nu}(\vec q_{1},\vec q_{2}),
\label{eq19}
\end{equation}
where $\nu=T,P$. The quantities $F^{C}_{\nu}$ and $F^{C(2)}_{\nu}$ are the model one- and two-body form factors which can be obtained from Eqs. (\ref{eq16}) and (\ref{eq17})  by replacing the intrinsic densities by the model ones, and $\theta_{\nu}$ is the c.m. correlation correction factor. Here it is worth mentioning that the above relations are exact when $\Phi_{C}$ is chosen to be the fully antisymmetric oscillator wave function. In this situation $\theta_{\nu}$ has the form
\begin{equation}
\theta_{\nu}(q)=\exp(q^{2}/4M_{\nu}\alpha^{2}_{\nu})
\label{eq20}
\end{equation}
where $\alpha^{2}_{\nu}$ is the oscillator constant and $M_{\nu}$ is the mass number of the nucleus. Moving further, it is assumed that the model two-body density may be written in terms of the model one-body density($\rho^{C}_{\nu}(r)$) as 
\begin{equation}
\rho^{C(2)}_{\nu}(\vec r_{1},\vec r_{2})= \rho^{C}_{\nu}(\vec r_{1})\rho^{C}_{\nu}(\vec r_{2}) \left[1-g( \vert \vec r_{1}-\vec r_{2} \vert )\right],  
\label{eq21}
\end{equation}
where  $g( \vert \vec r_{1}-\vec r_{2} \vert )$ is the phenomenological correlation function, satisfying the following requirements. (i) It has to be sufficiently short range, (ii) it should approach unity for $r \rightarrow 0$ to account for the hard core in the NN interaction, and (iii) its volume integral must be zero. The last condition ensures the normalization of $\rho^{C(2)}_{\nu}(\vec r_{1},\vec r_{2})$ so that its integral with respect to any one of its coordinates gives model one-body density. For more details about g(r) and its possible expressions, we refer the work of Ahmad \cite{18}.
 
Now using Eqs. (18), (19) and (21), the intrinsic two-body form factor may be written as \cite{18}
\begin{gather}
F^{(2)}_{\nu}(\vec q_{1},\vec q_{2})=\theta_{\nu}(\vec q_{1}+\vec q_{2})  
\Bigl[\frac{F_{\nu}(q_{1})F_{\nu}(q_{2})}{\theta_{\nu}(q_{1})\theta_{\nu}(q_{2})}~~~~~~~~~~~~~~~~~~~~~~~~~~\nonumber\\
~~~~~~~~~~~~~~~~~~~~~~~~~~~~~~~~~~-\tilde{g} \left( \frac{\vec q_{1}-\vec q_{2}}{2} \right)D^{C}_{\nu}(\vec q_{1}+\vec q_{2})\Bigr],
\label{eq22}	 
\end{gather}
where $\tilde{g}$(q) and $D^{C}_{\nu}$(q) are the Fourier transform of g(r) and $(\rho^{C}_{\nu}(r))^{2}$, respectively:
\begin{equation}
\tilde{g}(q)=\int e^{i\vec q.\vec r}g(r)d\vec r,
\label{eq23}
\end{equation}  
\begin{equation}
D^{C}_{\nu}(q)=\int e^{i\vec q.\vec r}(\rho^{C}_{\nu}(r))^{2}d\vec r.
\label{eq24}
\end{equation}
The model one-body density in Eq. (\ref{eq24}) may be obtained from the inverse Fourier transform of Eq. (\ref{eq18}), and takes the following form:
\begin{equation} 
\rho^{C}_{\nu}(r)=(1/2\pi^{2})\int j_{0}(qr)\frac{F_{\nu}(q)}{\theta_{\nu}(q)}q^{2}dq,
\label{eq25}
\end{equation}
where $j_{0}$ is the zeroth order spherical Bessel function.

\subsection{Nucleus-nucleus charge-changing cross section}

As we know, the reaction cross section ($\sigma_{R}$) (charge-changing cross section ($\sigma_{cc}$)) is defined as the total cross section for change in the mass (charge) number of the projectile. In a similar way, the neutron removal cross section ($\sigma_{-xn}$) can be defined as the cross section for the processes resulting in a change of the neutron number of the projectile. Thus, the total reaction cross section, $\sigma_{R}$, is the sum of $\sigma_{cc}$ and $\sigma_{-xn}$:
\begin{equation}
\sigma_{R}=\sigma_{cc}+\sigma_{-xn}
\label{eq26}
\end{equation} 
In the above relation, if we take $\sigma^{p}_{cc}$ as the contribution to the charge-changing cross section due to the scattering of only projectile protons, and represent the contribution of the rest of reaction, that takes care of the projectile neutron contribution to the charge-changing cross section ($\sigma^{n}_{cc}$) and the neutron removal cross section, by $\sigma^{rest}_{cc}$, the reaction cross section reads as
\begin{equation}
\sigma_{R}=\sigma^{p}_{cc}+\sigma^{rest}_{cc}.
\label{eq27}
\end{equation}
If $\eta$ provides an estimate to how much fraction of $\sigma^{rest}_{cc}$ is the projectile neutron contribution to the $\sigma_{cc}$, we may write
\begin{equation}
\sigma_{cc}=\sigma^{p}_{cc}+\eta~\sigma^{rest}_{cc},
\label{eq28}
\end{equation}
with
\begin{equation}
\sigma^{n}_{cc}=\eta ~ \sigma^{rest}_{cc}.
\label{eq29}
\end{equation}

As mentioned earlier, the projectile neutron contribution to the charge-changing cross section ($\sigma^{n}_{cc}$), within the framework of Glauber model, has been estimated following two approaches:
(i) The one is that of Bhagwat and Gambhir \cite{12}, in which $\sigma^{n}_{cc}$ is written in the same form as in Eq. (\ref{eq29}). It is clear that the evaluation of $\sigma^{n}_{cc}$ involves a parameter $\eta$ and requires both the proton and neutron densities of the projectile. Employing the RMF proton and neutron densities, and taking the parameter $\eta$=0.8$(Z_{P}/N_{P})^{2}$ for $N_{P}\geq Z_{P}$ and $\eta$=0.8 for $N_{P}\leq Z_{P}$, the authors \cite{12} are able to reproduce the charge-changing cross section date fairly well.
(ii) Whereas, in the second approach, Yamaguchi \textit{et al.} \cite{5} and Li \textit{et al.} \cite{17} introduce a phenomenological correction parameter that is responsible to include the effect of the presence of projectile neutrons to charge-changing cross section, and modifies the $\sigma^{p}_{cc}$ as follows:
\begin{equation}
\sigma_{cc}=\epsilon (E) ~ \sigma^{p}_{cc},
\label{eq30}
\end{equation} 
where $\epsilon (E)$ is the correction parameter which is defined as the ratio of the experimental $\sigma_{cc}$ and calculated $\sigma^{p}_{cc}$ values ($\sigma^{exp}_{cc}/\sigma^{p}_{cc}$). Contrary to the approach of Bhagwat and Gambhir \cite{12},  Yamaguchi \textit{et al.} \cite{5} and Li \textit{et al.} \cite{17} involve only the projectile protons in the calculations of $\sigma_{cc}$. Taking the correction parameter $\epsilon (E)$=1.107+0.01191$\times$$\exp$(1.444-0.004623E) \cite{17}, where E(in MeV) is the projectile energy/nucleon, and varying the projectile proton density parameter, Li \textit{et al.} \cite{17} have analysed the experimental charge-changing cross sections and predicted the root-mean-square proton radii for B, C, N, O, and F isotopes.

Comparison of the above mentioned approaches reveals that if we are interested to extract exclusively the projectile proton radius, it is reasonable to follow the approach of  Yamaguchi \textit{et al.} \cite{5} and Li \textit{et al.} \cite{17} as it involves only the projectile proton density in the analysis of charge-changing cross section data. However, we shall adopt a different prescription to get the value of the correction parameter $\epsilon (E)$, described in the next section. Keeping this in mind, let us now discuss the calculation of $\sigma^{p}_{cc}$ in the context of the present work. Following Eq. (\ref{eq1}), the $\sigma^{p}_{cc}$ is given by 
\begin{equation}
\sigma^{p}_{cc}=2\pi\int \left[1- \vert S^{p}_{el}(b)\vert^{2}\right]b~db,
\label{eq31} 
\end{equation}
where
\begin{equation}
S^{p}_{el}({b})\approx S^{p}_{0}(b)+S^{p}_{2}(b).
\label{eq32}
\end{equation}
The quantities $S^{p}_{0}(b)$ and $S^{p}_{2}(b)$ that consider only the projectile protons can be obtained by setting $N_{P}$=0 in the respective expressions for $S_{0}(b)$[Eq. (\ref{eq9})] and $S_{2}(b)$[Eq. (\ref{eq11})]. Such a consideration leads to the following expressions for $S^{p}_{0}(b)$ and $S^{p}_{2}(b)$:
\begin{equation}
S^{p}_{0}(b)= [1-\Gamma^{pp}_{00}(b)]^{Z_{P}Z_{T}}[1-\Gamma^{pn}_{00}(b)]^{Z_{P}N_{T}}, 
\label{eq33}
\end{equation}
\begin{gather}
S^{p}_{2}(b)=-\frac{A.Z_{P}}{8\pi^{2}k^{2}}(1-\Gamma^{NN}_{00}(b))^{A.Z_{P}-2}[(A-1)(Z_{P}-1)\nonumber\\
~~~~~~~~~~~~~~~~~\times (G_{22}(b)-G^{2}_{00}(b))+(Z_{P}-1)\nonumber\\
~~~~~~~~~~~~~~~~~\times (G_{21}(b)-G^{2}_{00}(b))+(A-1)~~\nonumber\\
~~~~\times (G_{12}(b)-G^{2}_{00}(b))],
\label{eq34}
\end{gather}
The quantities $G_{22}$, $G_{21}$, $G_{12}$, and $G_{00}$ in the above equation, though assume similar expressions (\ref{eq12})-(\ref{eq15}), but now they involve only the projectile proton density instead of projectile matter density.

\subsection{Nucleus-proton reaction and charge-changing cross sections}

Finally, the expressions for $S_{0}(b)$($S^{p}_{0}(b)$) and $S_{2}(b)$($S^{p}_{2}(b)$) can also be used to accommodate the scattering of a nucleus from a proton target. For this, we further set $Z_{T}=1$ and $N_{T}=0$. This simplification leads to the following expressions for $S_{0}(b)$($S^{p}_{0}(b)$) and $S_{2}(b)$($S^{p}_{2}(b)$) which can be used for calculating the reaction (charge-changing) cross section for nucleus-proton scattering:
\begin{equation}
S_{0}(b)= [1-\Gamma^{pp}_{0}(b)]^{Z_{P}}[1-\Gamma^{np}_{0}(b)]^{N_{P}}, 
\label{eq35}
\end{equation} 
\begin{equation}
S_{2}(b)=-\frac{B(B-1)}{8\pi^{2}k^{2}}[1-\Gamma^{NN}_{0}(b)]^{B-2}[G_{2}(b)-G^{2}_{0}(b)],
\label{eq36}
\end{equation}
\begin{equation}
S^{p}_{0}(b)= [1-\Gamma^{pp}_{0}(b)]^{Z_{P}}, 
\label{eq37}
\end{equation} 
\begin{equation}
S^{p}_{2}(b)=-\frac{Z_{P}(Z_{P}-1)}{8\pi^{2}k^{2}}[1-\Gamma^{NN}_{0}(b)]^{Z_{P}-2}[G_{2}(b)-G^{2}_{0}(b)],
\label{eq38}
\end{equation}
with
\begin{equation}
\Gamma^{ij}_{0}({b})=\int \rho^{i}_{P}(\vec r)\Gamma_{ij}(\vec b-\vec s)~d\vec r,
\label{eq39}
\end{equation}
\begin{gather}
G_{2}(b)=\int  \exp[-i(\vec q_{1}+\vec q_{2}).\vec b]F^{(2)}_{P}(\vec q_{1},\vec q_{2})~~~~~\nonumber\\ ~~~~~~~~\times f_{NN}(q_{1})f_{NN}(q_{2})d^{2}q_{1}d^{2}q_{2},
\label{eq40}
\end{gather} 
and
\begin{equation} 
\begin{aligned}
G_{0}(b) & =\int  e^{-i\vec q.\vec b}F_{P}(q)f_{NN}(q)d^{2}q, \\ 
& = (2\pi ik)\Gamma^{NN}_{0}(b).
\end{aligned}
\label{eq41}
\end{equation}
Moreover, it may be mentioned that $G_{2}(b)$ and $G_{0}(b)$ in expression (\ref{eq36}) involve the matter density of the projectile, whereas $G_{2}(b)$ and $G_{0}(b)$ in expression (\ref{eq38}) use only the projectile proton density.

\section{Results and Discussion}
\label{sec3}

Following the approach outlined in Sec. \ref{sec2}, we have analyzed the (i) charge-changing and interaction cross sections
for beryllium, boron, carbon, nitrogen, oxygen, and fluorine isotopes on a $^{12}$\rm C target, (ii) charge-changing cross sections for carbon and nitrogen isotopes on a proton target, and (iii) reaction cross sections for carbon isotopes on a proton target at medium energies. The inputs required in
the calculation are (i) the $NN$ scattering amplitude, and (ii) the proton and neutron density distributions of the nuclei under consideration. In connection with the interaction cross section ($\sigma_{I}$), one may note that, at energies under consideration, the contribution 
from inelastic scattering is negligible \cite{21}, $\sigma_{I}$ can be assumed to be nearly equal to the reaction cross section ($\sigma_{R}$). 
Hence, the Glauber model S-matrix ($S_{el}(b)$) in Eq. (\ref{eq2}) can be used to calculate both $\sigma_{I}$ as well as $\sigma_{R}$ from Eq. (\ref{eq1}). 

The $NN$ scattering amplitude takes care of the nuclear in-medium effects, arising due to phase variation, higher momentum transfer components,
and Pauli blocking. For this, we have parameterized the $NN$ scattering amplitude as follows \cite{22}:
\begin{gather}
f_{NN}(\vec q)=\biggl[\frac{ik\sigma_{NN}}{4\pi}\sum^{\infty}_{n=0}A_{n+1}\left(\frac{\sigma_{NN}}
{4\pi\beta_{NN}}\right) ^{n}\frac{(1-i\rho_{NN})^{n+1}}{(n+1)}\nonumber\\
~~~~~~~~~~~~\times \exp\left(\frac{-\beta_{NN}q^{2}}{2(n+1)}\right)\biggr]\exp\left(\frac{-i\gamma_{NN} q^{2}}{2}\right),
\label{eq42}
\end{gather}
where
\begin{gather}
A_{n+1}=\frac{A_{1}}{n(n+1)}+\frac{A_{2}}{(n-1)n}~~~~~~~~~~~~~~~~\nonumber\\
~~~~~~~~~+\frac{A_{3}}{(n-2)(n-1)}+.....+\frac{A_{n}}{1.2}~,
\label{eq43}
\end{gather}
with $A_{1}=1$.\\

The $NN$ amplitude [Eq. (\ref{eq42})] consists of four adjustable parameters; 
$\sigma_{NN}$, $\rho_{NN}$, $\beta_{NN}$, and $\gamma_{NN}$. The values of these parameters have been obtained in our earlier work \cite{11} from the 
simultaneous description of the in-medium $NN$ total cross section \cite{23} and the proton-nucleus elastic differential cross section data at 650, 800,
and 1000 MeV which cover the  required energy range in this work; the in-medium values of the $NN$ amplitude parameters at the desired energies are
obtained by a linear interpolation/extrapolation of their values at 650, 800, and 1000 MeV. 

The intrinsic proton and neutron densities of the projectile are obtained from the Slater determinants consisting of the harmonic 
oscillator single-particle wave functions \cite{11}. These densities 
involve the oscillator constant as their basic input, which assumes different values for different nuclei and are obtained from the analysis 
of charge-changing and interaction cross sections in the present work. The oscillator constants for proton ($\rm \alpha^{2}_{p}$) and neutron 
($\rm \alpha^{2}_{n}$) density distributions are related to the corresponding rms proton ($\rm \langle r^2_{p}\rangle^{1/2}$) and neutron ($\rm \langle r^2_{n}\rangle^{1/2}$) 
radii of a given nucleus. For the target $^{12}$\rm C nucleus, we involve the charge density as obtained from the electron scattering experiment \cite{24} and assume the neutron and proton densities to be the same. Here, it is important to note that the intrinsic proton and neutron densities simulate the finite size of the nucleon through free variation of the oscillator constant, whose values, as pointed out above, are adjusted from the study of experimental charge-changing and interaction cross sections. As a result, the rms proton radius, obtained in this work, gives directly the charge radius ($\rm \langle r^2_{ch}\rangle^{1/2}$) of the nucleus, and the corresponding neutron radius also takes care of the finite size of the nucleon. In the following, we refer our deduced rms proton(neutron) radius as $r_{p}$$(r_{n})$, and the corresponding rms point proton(neutron) radius will be referred to as $r_{p}^{pt}$$(r_{n}^{pt})$.
 
\subsection{Nucleus-nucleus charge-changing cross section}

As mentioned in Sec.\ref{sec2}B, the calculation of charge-changing cross section ($\sigma_{cc}$) requires a phenomenological correction parameter
$\epsilon(E)$ to account for the contribution due to the presence of neutrons in the projectile. Also, it has been pointed out that in one
calculation \cite{17}, $\epsilon(E)$ though assumes energy dependent parameterization but its value attains almost a constant value beyond 600
MeV/nucleon, and was taken to be the same for all the isotopes of different elements. Whereas in some other calculation \cite{12}, $\epsilon(E)$ assumes isotopic dependent parametrization without reference to 
the energy of the projectile, giving rise different values of the correction parameter for all the isotopes of different elements. In this work, we adopt the former approach \cite{17}, but proceed to find $\epsilon(E)$ in a different way. To start with, we involve only those projectile nuclei for which the experimental charge radii \cite{25,26} as well as their experimental charge-changing cross sections are available. For such nuclei, we adjust the oscillator constant, $\rm \alpha^{2}_{p}$, to get their proton densities that lead to the experimentally known charge radii ($\rm \langle r^2_{ch}\rangle^{1/2}$). Then, with these proton distributions, we compute the contribution to CCCS due to the projectile protons, $\sigma^{p}_{cc}$, using Eqs. (\ref{eq31})-(\ref{eq34}), and find the value of $\epsilon(E)$ (=$\sigma^{exp}_{cc}/\sigma^{p}_{cc}$) in each case. The results of such calculations with a $^{12}$\rm C target are presented in Table \ref{tab1}. It is found that the correction parameter, $\epsilon(E)$, within the isotopic chain 
of a given element is nearly the same and energy independent, and its average value (given in the last column of Table \ref{tab1}) shows some variation as we 
move on to different elements. In the case of fluorine, the experimental charge radius is available
for $^{19}$\rm F only, the correction parameter obtained for it is thus used for rest of its isotopes. Because of the energy independence of correction parameter, we, hereafter, represent it as $\epsilon$ instead of $\epsilon(E)$. 

To test the suitability of the present method used to extract the correction parameter $\epsilon$, it is worthwhile to revisit the nuclei, except for $^{19}$\rm F, in Table \ref{tab1} and obtain the required oscillator constant for proton distribution ($\alpha^{2}_{p}$) and the corresponding proton radius ($r_{p}$), that may now reproduce $\sigma^{exp}_{cc}$ using the respective (average) value of $\epsilon$. In this way, the difference between the extracted value of $r_{p}$ and the corresponding experimental one  ($\rm \langle r^2_{ch}\rangle^{1/2}$) will provide an estimate of the uncertainty of the present approach. The results of such calculations are presented in Table \ref{tab2}. In this table, $\alpha^{2}_{p}$ and $r_{p}$ provide, respectively, the values of the oscillator constant and the corresponding proton radius that now reproduces the experimental CCCS using the average value of the correction parameter $\epsilon$. The last coulmn in Table \ref{tab2} gives the percentage difference between the extracted and experimental proton radii. It is found that the maximum deviation is about 2\%. Thus, it seems that the present method for obtaining the correction parameter $\epsilon$ is quite reliable, and only about not more than 2\% uncertainty may be involved in predicting the proton radii of those exotic isotopes whose experimental charge radii are not known.             

\begin{figure}
	\begin{center}
		\includegraphics[height=10.5cm, width=8.6cm]{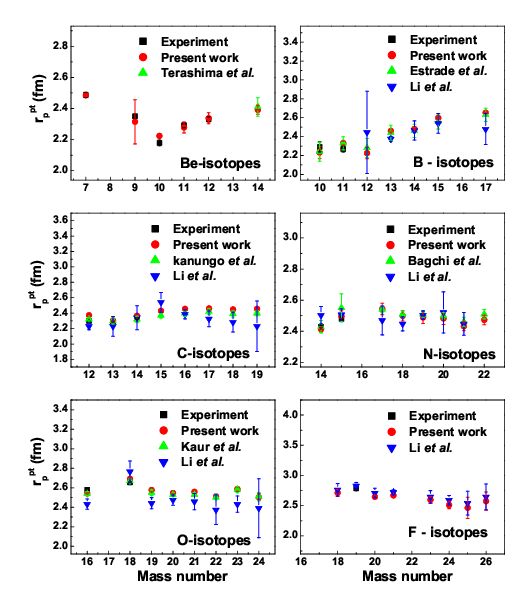}
		\caption{The point proton radii corresponding to our deduced  proton radii  given in Tables \ref{tab2} and \ref{tab3}. The results of other works are  presented for comparison. The available experimental values of point proton radii (filled squares) are obtained from  Refs.  \cite{25} and \cite{26} after correcting for finite size of the nucleon . The values denoted by Terashima  \textit{et al.}, Estrade \textit{et al.}, Li \textit{et al.}, Kanungo \textit{et al.}, Bagchi \textit{et al.}, and Kaur \textit{et al.} are taken from Refs. \cite{7}, \cite{6}, \cite{17}, \cite{8}, \cite{9}, and \cite{10}, respectively.}
		\label{fig1}
		\end{center}
\end{figure}

Having discussed the procedure for obtaining the correction parameter $\epsilon$, needed to incorporate the effect of the presence of projectile neutrons in 
calculating the CCCS, we now proceed to reproduce the experimental CCCS data for $^{14}$\rm Be, $^{12-15,17}$\rm B, $^{15-19}$\rm C,
$^{17-22}$\rm N, $^{19-24}$\rm O, and $^{18,20,21,23-26}$\rm F isotopes on a $^{12}$\rm C target by adjusting the oscillator constant in SDHO proton densities and using the respective average value of $\epsilon$. The results of 
these calculations are given in Table \ref{tab3}. Like in Table \ref{tab2}, $\alpha^{2}_{p}$ and $r_{p}$ in Table \ref{tab3} give, respectively, values of the oscillator constant and the corresponding proton radius that can reproduce the experimental CCCS for the above mentioned isotopes. It is noticed that the gradual filling of neutrons in the $1d_{5/2}$ orbital reflects a consistent decrease in the proton radii up to N = 14 for both the nitrogen and oxygen isotopes. At N = 14, we find a local minimum which is an indication for this new subshell closure. Similar result has also been reported by Bagchi \textit{et al.} \cite{9} and Kaur \textit{et al.} \cite{10} for nitrogen and oxygen isotopes, respectively. Thus, our findings further support the claim \cite{9,10} that the possible origin of the subshell closure at N = 14 is the attractive isospin (T) = 0 tensor interaction between the protons in the $1p_{1/2}$ orbital and neutrons in the $1d_{5/2}$ orbital, which reduces the gap between proton $1p_{1/2}$ and $1p_{3/2}$ orbitals. As a result, the $1p_{1/2}$ orbital gets lowered, leading to a dip in the proton radius at N = 14. In Fig. \ref{fig1}, we compare our predicted point proton radii ($r_{p}^{pt}$), obtained from the proton radii (Tables \ref{tab2} and \ref{tab3}) after correcting for finite size of the nucleon, with some of the reasults available in the literature; For completeness, we have also included in Fig. \ref{fig1} the experimental point proton radii corresponding to charge radii of the nuclei \cite{24,25} given in Table \ref{tab1}.  It is found that our extracted point proton radii agree fairly well in all the cases. Thus, the present approach seems to be quite justified for the study of CCCS, and the extracted proton radii of exotic neutron rich nuclei can be considered as reliable estimates for onward calculations.

\begin{longtable*}{cccccccc}
 \caption{The oscillator constant, $\rm \alpha_p^2$, gives the experimental projectile charge radius 
 	($\rm \langle r^2_{ch}\rangle^{1/2}$) \cite{24,25}. $\sigma_{CC}^{p}$ provides the contribution to charge-changing cross section due to projectile protons at energy E, taking $^{12}$\rm C as a target. The correction parameter
 $\rm \epsilon \left (=\sigma_{CC}^{exp}/\sigma_{CC}^{p} \right )$ is the ratio of the experimental charge-changing cross section ($\rm \sigma_{CC}^{exp}$) 
 and $\rm \sigma_{CC}^{p}$. The last column gives the average value of $\rm \epsilon$, $\rm \epsilon_{avg}$, for the considered isotopes of a given element.}
 \label{tab1}
\\
\hline
\\
$\rm Projectile$    &$\rm E/A$ &$\rm \alpha_p^2$    &$\rm \langle r^{2}_{ch}\rangle^{1/2}$      &$\rm \sigma_{CC}^{p}$ &$\rm \sigma_{CC}^{exp}$  &$\rm \epsilon (=\sigma_{CC}^{exp}/\sigma_{CC}^{p})$ &$\rm \epsilon_{avg}$ \\
                &$\rm (MeV)$  &$\rm (fm^{-2})$     &\rm (fm)                               &$\rm (mb)$              &$\rm (mb)$               &                                                                            &                     \\
                 &                    &                           &              &       &                         &                                                                            &                     \\

\hline
\\
$\rm ^{7}Be$     &772 &0.2549$\pm$0.0031   &2.6468$\pm$0.0161\cite{25}                             &658.2$\pm$2.46           &706$\pm$8\cite{7}        &1.073$\pm$0.008         &1.073$\pm$0.013  \\
$\rm ^{9}Be$      &921 &0.2889$\pm$0.0027   &2.5190$\pm$0.0120\cite{25}                             &640.3$\pm$1.79           &682$\pm$30\cite{7}       &1.065$\pm$0.044         &                   \\
$\rm ^{10}Be$    &946  &0.3318$\pm$0.0046   &2.3612$\pm$0.0166\cite{25}                             &618.1$\pm$2.49           &670$\pm$10\cite{7}       &1.084$\pm$0.011         &                   \\
$\rm ^{11}Be$   &962 &0.3062$\pm$0.0036   &2.4669$\pm$0.0147\cite{25}                              &636.7$\pm$2.23           &681$\pm$3 \cite{7}       &1.069$\pm$0.001         &                   \\
$\rm ^{12}Be$    &925 &0.2993$\pm$0.0037   &2.5031$\pm$0.0157\cite{25}                             &638.2$\pm$2.38           &686$\pm$3 \cite{7}       &1.075$\pm$0.001         &                   \\
\\
\hline
\\
$\rm ^{10}B$    &925 &0.3309$\pm$0.0132   &2.4277$\pm$0.0499\cite{24}                              &680.2$\pm$8.7           &685$\pm$14\cite{6}       &1.007$\pm$0.008         &1.023$\pm$0.005  \\
$\rm ^{11}B$    &932 &0.3392$\pm$0.0081   &2.4060$\pm$0.0294\cite{24}                              &676.0$\pm$5.1           &702$\pm$6\cite{6}        &1.038$\pm$0.001         &                   \\
\\
\hline
\\
$\rm ^{12}C$    &937  &0.3346$\pm$0.0006   &2.4702$\pm$0.0022\cite{24}                             &731.4$\pm$0.4           &733$\pm$7\cite{8}       &1.002$\pm$0.009         &0.993$\pm$0.008  \\
$\rm ^{13}C$     &828  &0.3386$\pm$0.0009   &2.4614$\pm$0.0034\cite{24}                            &736.0$\pm$0.7           &726$\pm$7\cite{8}       &0.986$\pm$0.009         &                   \\
$\rm ^{14}C$    &900 &0.3289$\pm$0.0023   &2.5025$\pm$0.0087\cite{24}                              &737.4$\pm$1.7           &731$\pm$7\cite{8}       &0.991$\pm$0.007         &                   \\
\\
\hline
\\
$\rm ^{14}N$    &932    &0.3219$\pm$0.0017   &2.5582$\pm$0.0070\cite{24}                            &785.8$\pm$1.5           &793$\pm$9\cite{9}       &1.009$\pm$0.009         &1.014$\pm$0.016  \\
$\rm ^{15}N$     &776 &0.3114$\pm$0.0019   &2.6058$\pm$0.0080\cite{24}                             &801.8$\pm$1.7           &816$\pm$20\cite{9}       &1.018$\pm$0.023         &                   \\
\\
\hline
\\
$\rm ^{16}O$   &857  &0.2959$\pm$0.0011   &2.6991$\pm$0.0052\cite{24}                              &854.4$\pm$1.1           &848$\pm$4\cite{10}       &0.992$\pm$0.003         &1.002$\pm$0.004  \\
$\rm ^{18}O$    &872  &0.2819$\pm$0.0011   &2.7726$\pm$0.0056\cite{24}                             &869.6$\pm$1.3           &879$\pm$5\cite{10}       &1.011$\pm$0.004         &                   \\
\\
\hline
\\
$\rm ^{19}F$   &930  &0.2751$\pm$0.0005   &2.8976$\pm$0.0025\cite{9}                              &928.9.3$\pm$0.7           &1016$\pm$10\cite{4}      &1.094$\pm$0.010         &1.094$\pm$0.010  \\
\\
\hline

\end{longtable*}

\begin{longtable*}{ccccccccc}
	\caption{Revisit of the projectiles in Table I: The oscillator constant, $\rm \alpha_p^2$, gives the projectile proton radius 
		($r_{p}$) obtained by fitting the experimental charge-changing cross section, $\rm \sigma_{CC}^{exp}$, at energy E, using the average value of the correction parameter, $\rm \epsilon_{avg}$, given in Table \ref{tab1}. The $(\Delta\rm r_{p})\%$ gives the percentage difference between the experimental projectile charge radius ($\rm \langle r^2_{ch}\rangle^{1/2}$) and $r_{p}$.}
\label{tab2}
\\
\hline
\\
$\rm Projectile$   &$\rm E/A$ &$\rm \alpha_p^2$   &$r_{p}$ &$\rm \sigma_{CC}^{exp}$         &$(\Delta\rm r_{p})\%$   \\
&$\rm (MeV)$ &$\rm (fm^{-2})$     &$\rm (fm)$                               &$\rm (mb)$              &$\rm (fm)$                                                                                                          \\
	&                    &                           &$\rm (This~work)$                     &                   &$\rm$                                                                                               
\\

\hline
\\
$\rm ^{7}Be$   &772  &0.2554$\pm$0.0006   &2.6444$\pm$0.0029                                        &706$\pm$8\cite{7}        &0.09          \\
$\rm ^{9}Be$   &921  &0.2965$\pm$0.0293   &2.4865$\pm$0.1328                                        &682$\pm$30\cite{7}       &1.29                           \\
$\rm ^{10}Be$   &946 &0.3205$\pm$0.0030   &2.4025$\pm$0.0113                                        &670$\pm$10\cite{7}       &1.74                            \\
$\rm ^{11}Be$   &962 &0.3097$\pm$0.0083   &2.4530$\pm$0.0323                                         &681$\pm$3 \cite{7}       &0.56                            \\
$\rm ^{12}Be$  &925   &0.2978$\pm$0.0079   &2.5094$\pm$0.0326                                        &686$\pm$3 \cite{7}       &0.25                            \\
\\
\hline
\\
$\rm ^{10}B$   &925  &0.3470$\pm$0.0171   &2.3704$\pm$0.0606                                         &685$\pm$14\cite{6}           &2.36  \\
$\rm ^{11}B$   &932  &0.3230$\pm$0.0052   &2.4655$\pm$0.0201                                        &702$\pm$6\cite{6}        &2.47                            \\
\\
\hline
\\
$\rm ^{12}C$    &937  &0.3255$\pm$0.0011   &2.5045$\pm$0.0041                                        &733$\pm$7\cite{8}       &1.38          \\
$\rm ^{13}C$    &828 &0.3455$\pm$0.0009   &2.4367$\pm$0.0033                                         &726$\pm$7\cite{8}       &1.00                            \\
$\rm ^{14}C$    &900  &0.3306$\pm$0.0013   &2.4961$\pm$0.0047                                        &731$\pm$7\cite{8}       &0.25                            \\
\\
\hline
\\
$\rm ^{14}N$     &932  &0.3260$\pm$0.0041   &2.5424$\pm$0.0159                                        &793$\pm$9\cite{9}       &0.61           \\
$\rm ^{15}N$    &776  &0.3074$\pm$0.0073   &2.6227$\pm$0.0318                                        &816$\pm$20\cite{9}       &0.64                            \\
\\
\hline
\\
$\rm ^{16}O$    &857  &0.3036$\pm$0.0011   &2.6651$\pm$0.0049                                        &848$\pm$4\cite{10}       &1.25          \\
$\rm ^{18}O$   &872  &0.2749$\pm$0.0011   &2.8077$\pm$0.0058                                         &879$\pm$5\cite{10}       &1.26                            \\
\\
\hline
\\

\end{longtable*}

\begin{longtable*}{ccccccccc}
 \caption{The oscillator constant, $\rm \alpha_p^2$, gives the projectile proton radius 
 	($r_{p}$) obtained by fitting the experimental charge-changing cross section, $\rm \sigma_{CC}^{exp}$, at energy E with $^{12}$\rm C as a target, using the average value of the correction parameter, $\rm \epsilon_{avg}$, given in Table \ref{tab1}.}
 \label{tab3}
\\
\hline
\\
$\rm Projectile$  &\rm E/A    &$\rm \alpha_p^2$     &$r_{p}$ &$\rm \sigma_{CC}^{exp}$   \\
               &\rm (MeV)  &$\rm (fm^{-2})$      &$\rm (fm)$                          &$\rm (mb)$                   \\
               &           &                     & $\rm (This~work)$                                                                                                                        \\

\hline
\\
$\rm ^{14}Be$  &833        &0.2886$\pm$0.0062    &2.5612$\pm$0.0271                   &697$\pm$4\cite{7}   \\
\\
\hline
\\
$\rm ^{12}B$   &991        &0.3529$\pm$0.0162    &2.3657$\pm$0.0561                   &691$\pm$13\cite{6}  \\
$\rm ^{13}B$   &897        &0.2965$\pm$0.0029    &2.5872$\pm$0.0127                   &723$\pm$6\cite{6}   \\
$\rm ^{14}B$   &926        &0.2930$\pm$0.0004    &2.6080$\pm$0.0018                   &727$\pm$4\cite{6}   \\
$\rm ^{15}B$   &920        &0.2707$\pm$0.0018    &2.7181$\pm$0.0091                   &747$\pm$5\cite{6}   \\
$\rm ^{17}B$   &862        &0.2618$\pm$0.0009    &2.7721$\pm$0.0051                   &759$\pm$4\cite{6}   \\
\\
\hline
\\
$\rm ^{15}C$   &907        &0.3156$\pm$0.0006    &2.5589$\pm$0.0024                   &743$\pm$7\cite{8}  \\
$\rm ^{16}C$   &907        &0.3109$\pm$0.0009    &2.5821$\pm$0.0038                   &748$\pm$7\cite{8}  \\
$\rm ^{17}C$   &979        &0.3101$\pm$0.0007    &2.5889$\pm$0.0029                   &754$\pm$7\cite{8}  \\ 
$\rm ^{18}C$   &895        &0.3143$\pm$0.0013    &2.5746$\pm$0.0053                   &747$\pm$7\cite{8}  \\
$\rm ^{19}C$   &895        &0.3128$\pm$0.0035    &2.5833$\pm$0.0147                   &749$\pm$9\cite{8}  \\
\\
\hline
\\
$\rm ^{17}N$   &938        &0.2990$\pm$0.0081    &2.6664$\pm$0.0354                   &819$\pm$5\cite{9}  \\
$\rm ^{18}N$   &927        &0.3095$\pm$0.0077    &2.6238$\pm$0.0321                   &810$\pm$6\cite{9}  \\
$\rm ^{19}N$   &896        &0.3124$\pm$0.0089    &2.6143$\pm$0.0363                   &809$\pm$5\cite{9}  \\
$\rm ^{20}N$   &891        &0.3145$\pm$0.0091    &2.6080$\pm$0.0370                   &808$\pm$5\cite{9}  \\
$\rm ^{21}N$   &876        &0.3268$\pm$0.0070    &2.5606$\pm$0.0271                   &799$\pm$7\cite{9}  \\
$\rm ^{22}N$   &851        &0.3177$\pm$0.0072    &2.5993$\pm$0.0290                   &810$\pm$7\cite{9} \\
\\
\hline
\\
$\rm ^{19}O$   &956        &0.2983$\pm$0.0038    &2.6977$\pm$0.0174                   &852$\pm$7\cite{10}  \\
$\rm ^{20}O$   &880        &0.3056$\pm$0.0007    &2.6677$\pm$0.0032                   &846$\pm$4\cite{10}  \\
$\rm ^{21}O$   &937        &0.3033$\pm$0.0027    &2.6800$\pm$0.0121                   &847$\pm$6\cite{10}  \\
$\rm ^{22}O$   &937        &0.3143$\pm$0.0000    &2.6346$\pm$0.0000                   &837$\pm$3\cite{10}  \\
$\rm ^{23}O$   &871        &0.2980$\pm$0.0045    &2.7076$\pm$0.0205                   &857$\pm$8\cite{10}  \\
$\rm ^{24}O$   &866        &0.3168$\pm$0.0082    &2.6278$\pm$0.0344                   &839$\pm$11\cite{10}  \\
\\
\hline
\\
$\rm ^{18}F$   &930        &0.2880$\pm$0.0115    &2.8294$\pm$0.0582                   &998$\pm$25\cite{4}  \\
$\rm ^{20}F$   &930        &0.3029$\pm$0.0035    &2.7639$\pm$0.0161                   &980$\pm$13\cite{4}  \\
$\rm ^{21}F$   &930        &0.2979$\pm$0.0005    &2.7891$\pm$0.0024                   &986$\pm$10\cite{4}  \\
$\rm ^{23}F$   &930        &0.3146$\pm$0.0111    &2.7177$\pm$0.0493                   &967$\pm$22\cite{4}  \\
$\rm ^{24}F$   &930        &0.3346$\pm$0.0141    &2.6368$\pm$0.0574                   &946$\pm$24\cite{4}  \\
$\rm ^{25}F$   &930        &0.3468$\pm$0.0409    &2.5914$\pm$0.1678                   &934$\pm$54\cite{4}  \\
$\rm ^{26}F$   &930        &0.3207$\pm$0.0317    &2.6961$\pm$0.1440                   &962$\pm$48\cite{4}  \\

\\
\hline
\\
\end{longtable*}

\begin{figure}
	\begin{center}
		\includegraphics[height=10.5cm, width=8.6cm]{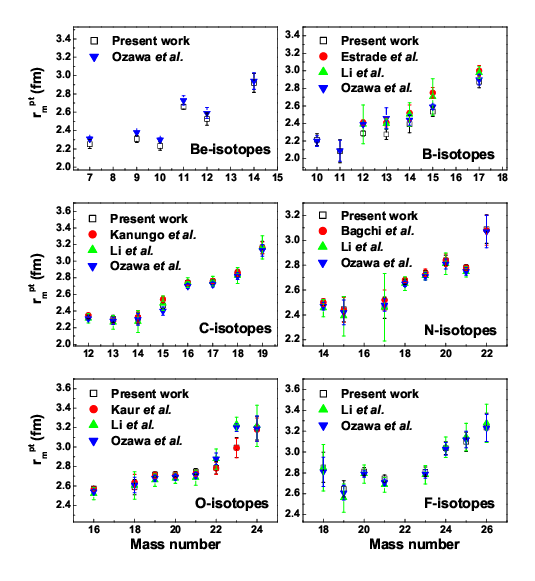}
		\caption{The point matter radii corresponding to our deduced proton (Tables \ref{tab2} and \ref{tab3}) and neutron (Table \ref{tab4}) radii. The results of other works are presented for comparison. The values denoted by Ozawa  \textit{et al.}, Estrade \textit{et al.}, Li \textit{et al.}, Kanungo \textit{et al.}, Bagchi \textit{et al.}, and Kaur \textit{et al.} are taken from Refs. \cite{31}, \cite{6}, \cite{17}, \cite{8}, \cite{9}, and \cite{10}, respectively.}
		\label{fig2}
	\end{center}
\end{figure}

\subsection{Nucleus-nucleus interaction cross section}

Our next goal is to study the interaction cross sections and extract the neutron (matter) radii of beryllium, boron, carbon, nitrogen, oxygen, and fluorine isotopes. 
Keeping the similar values of the oscillator constant for SDHO proton densities ($\rm \alpha^{2}_{p}$) as obtained from the analysis of charge-changing cross
sections (Tables \ref{tab2} and \ref{tab3}) on a $^{12}$\rm C target, we now adjust the oscillator constant for SDHO neutron densities ($\rm \alpha^{2}_{n}$) to reproduce the corresponding interaction cross sections.
Table \ref{tab4} provides our extracted neutron radii ($r_{n}$) of the nuclei under consideration. In Fig. \ref{fig2}, the (point) matter radii ($r_{m}^{pt}$), obtained using $r_{p}$ and $r_{n}$ of Tables \ref{tab2}, \ref{tab3} and \ref{tab4} after correcting for finite nucleon size, are compared with some earlier results. A fairly good agreement is seen in all the cases. Moreover, it may be added that the matter radii, as obtained in this work, though supplement the results of our earlier work \cite{11}, the results of the present calculations can be considered to be more reliable possibly due to fairly good estimates of the proton and neutron radii obtained, respectively, by a systematic analysis of charge-changing and interaction cross sections. 

\begin{figure}
	\begin{center}
		\includegraphics[scale=1.05]{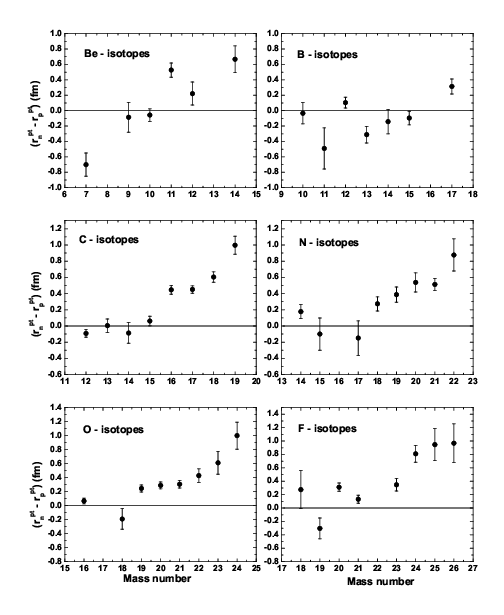}
		\caption{The neutron skin thickness as a function of mass number for Be, B, C, N, O, and F isotopes, using the point proton and point neutron radii corresponding to our deduced proton and neutron radii given in Tables \ref{tab2}, \ref{tab3}, and \ref{tab4}.}
		\label{fig3}
	\end{center}
\end{figure}

Next, we proceed to examine our extracted proton, neutron, and matter radii in the light of getting information about whether a given exotic neutron rich nucleus has a thick neutron skin or leads to a halo-like structure. As we know, the one-neutron ($S_{n}$) or two-neutron ($S_{2n}$) separation energy plays an essential role in distinguishing between a thick neutron skin and the existence of a halo structure in exotic neutron rich nuclei. Obviously, the smaller value of one- or two-neutron separation energy indicates a significant probability to find one neutron or two neutrons at large radius, and forms the basis of halo-like structure in which the nucleon density is extended to a large distance. Unfortunately, the SDHO densities, as obtained in the present work, cannot be justified to get the extended matter distribution in exotic nuclei, but we still expect that the effect of extended part of the neutron distribution may be simulated in our estimates of the matter radii for such nuclei \cite{27}. Keeping this in mind, we, therefore, consider the neutron skin thickness, $(r_{n}^{pt}-r_{p}^{pt})$, and study its behaviour along with the values of one-/two-neutron separation energies \cite{28}. Fig. \ref{fig3} depicts our deduced neutron skins for different isotopic chains as a function of the mass number. From this figure, we observe that within the isotopic chain of a given element, the nuclei $^{11}$\rm Be, $^{19}$\rm C, $^{22}$\rm N, and $^{26}$\rm F correspond to the largest neutron skin thickness, but with the minimum value of one-neutron separation energy, whereas the nuclei $^{14}$\rm Be and $^{17}$\rm B also correspond to the largest neutron skin thickness, but involve minimum value of two-neutron separation energy. This shows that the nuclei $^{11}$\rm Be, $^{19}$\rm C, $^{22}$\rm N, and $^{26}$\rm F exhibit one-neutron halo-like structure, whereas for $^{14}$\rm Be and $^{17}$\rm B two-neutron halo-like structure develop. To strengthen our information relating the halo-like behaviour of the above mentioned nuclei, we refer readers to see some earlier works \cite{8,9,10,29,30}. For rest of the neutron rich isotopes, having higher values of $S_{n}$/$S_{2n}$, the order of the thickness of neutron surface can be assessed directly from the measure of $(r_{n}^{pt}-r_{p}^{pt})$. For example, the data in Fig. \ref{fig3} reveal a thick neutron surface for $^{16-18}$\rm C, $^{20,21}$\rm N, $^{22-24}$\rm O, and $^{24,25}$\rm F isotopes.


\begin{table}
\begin{center}	       
 \caption{The oscillator constant, $\rm \alpha_n^2$, gives the projectile neutron radius 
 	($r_{n}$) obtained by fitting the experimental interaction cross section, $\rm \sigma_{I}^{exp}$, at energy E with $\rm ^{12}C$ as a target, keeping the value of oscillator constant in (SDHO) proton density distribution ($\rm \alpha_p^2$) the same as obtained in Tables \ref{tab2} and \ref{tab3}. The values of $\rm \sigma_{I}^{exp}$ for $^{22,23}\rm O$, with superscript $a$, are taken from Ref. \cite{32}}.
 \label{tab4}
 \begin{tabular}{ccccccccc}

\hline
\\
$\rm Projectile$  &$\rm E/A$    &$\rm \alpha_n^2$         &$r_{n}$  &$\rm \sigma_I^{exp}$ \\
               &$\rm (MeV)$  &$\rm (fm^{-2})$          &$\rm (fm)$  &\rm (mb)  \\
               &             &                         & $\rm (This~work)$                  &\cite{31}          \\
               &             &                         & \\      
\hline
\\
$\rm ^{7}Be$   &790          &0.4050$\pm$0.0490        &1.9994$\pm$0.1332  &738$\pm$9 \\
$\rm ^{9}Be$   &790          &0.3340$\pm$0.0138        &2.4059$\pm$0.0482  &806$\pm$9 \\
$\rm ^{10}Be$  &790          &0.3655$\pm$0.0191        &2.3489$\pm$0.0639  &813$\pm$10 \\
$\rm ^{11}Be$  &790          &0.2391$\pm$0.0087        &2.9479$\pm$0.0552  &942$\pm$8   \\
$\rm ^{12}Be$  &790          &0.2879$\pm$0.0216        &2.7168$\pm$0.1080  &927$\pm$18   \\
$\rm ^{14}Be$  &800          &0.2346$\pm$0.0193        &3.1937$\pm$0.1401  &1082$\pm$34   \\
\\
\hline
\\
$\rm ^{10}B$   &960          &0.3564$\pm$0.0194        &2.3391$\pm$0.0664  &789$\pm$16  \\
$\rm ^{11}B$   &950          &0.4989$\pm$0.0944        &2.0174$\pm$0.2230  &778$\pm$30  \\
$\rm ^{12}B$   &790          &0.3446$\pm$0.0023        &2.4623$\pm$0.0083  &866$\pm$7 \\
$\rm ^{13}B$   &790          &0.4045$\pm$0.0289        &2.2972$\pm$0.0867  &883$\pm$14 \\
$\rm ^{14}B$   &790          &0.3726$\pm$0.0406        &2.4746$\pm$0.1470  &929$\pm$26  \\
$\rm ^{15}B$   &740          &0.3475$\pm$0.0187        &2.6280$\pm$0.0737  &965$\pm$15\\
$\rm ^{17}B$   &800          &0.2737$\pm$0.0152        &3.0693$\pm$0.0890  &1118$\pm$22 \\
\\
\hline
\\
$\rm ^{12}C$   &950          &0.3487$\pm$0.0110        &2.4197$\pm$0.0391  &853$\pm$6 \\
$\rm ^{13}C$   &960          &0.3525$\pm$0.0204        &2.4401$\pm$0.0739  &862$\pm$12\\
$\rm ^{14}C$   &965          &0.3672$\pm$0.0323        &2.4157$\pm$0.1138  &880$\pm$19\\
$\rm ^{15}C$   &740          &0.3345$\pm$0.0133        &2.6159$\pm$0.0536  &945$\pm$10\\
$\rm ^{16}C$   &960          &0.2666$\pm$0.0082        &3.0043$\pm$0.0473  &1036$\pm$11\\
$\rm ^{17}C$   &965          &0.2753$\pm$0.0072        &3.0151$\pm$0.0402  &1056$\pm$10\\
$\rm ^{18}C$   &955          &0.2608$\pm$0.0090        &3.1473$\pm$0.0557  &1104$\pm$15\\             
$\rm ^{19}C$   &960          &0.2124$\pm$0.0107        &3.5334$\pm$0.0925  &1231$\pm$28\\
\\
\hline
\\
$\rm ^{14}N$   &965          &0.2874$\pm$0.0134        &2.7077$\pm$0.0654  &932$\pm$9\\
$\rm ^{15}N$   &975          &0.3362$\pm$0.0376        &2.5288$\pm$0.1544  &930$\pm$30    \\
$\rm ^{17}N$   &710          &0.3778$\pm$0.0450        &2.5266$\pm$0.1654  &965$\pm$24\\
$\rm ^{18}N$   &1020         &0.3021$\pm$0.0104        &2.8809$\pm$0.0509  &1046$\pm$8 \\
$\rm ^{19}N$   &1005         &0.2914$\pm$0.0100        &2.9800$\pm$0.0525  &1076$\pm$9  \\
$\rm ^{20}N$   &950          &0.2732$\pm$0.0127        &3.1178$\pm$0.0751  &1121$\pm$17   \\
$\rm ^{21}N$   &1005         &0.2926$\pm$0.0080        &3.0457$\pm$0.0425  &1114$\pm$9         \\
$\rm ^{22}N$   &965          &0.2346$\pm$0.0207        &3.4332$\pm$0.1618  &1245$\pm$49      \\
\\
\hline
\\
$\rm ^{16}O$   &970          &0.2898$\pm$0.0068        &2.7277$\pm$0.0326  &982$\pm$6\\
$\rm ^{18}O$   &1050         &0.3499$\pm$0.0330        &2.6281$\pm$0.1334  &1032$\pm$26\\
$\rm ^{19}O$   &970          &0.2927$\pm$0.0060        &2.9295$\pm$0.0305  &1066$\pm$9 \\
$\rm ^{20}O$   &950          &0.2996$\pm$0.0086        &2.9412$\pm$0.0431  &1078$\pm$10 \\
$\rm ^{21}O$   &980          &0.3015$\pm$0.0077        &2.9699$\pm$0.0387  &1098$\pm$11\\
$\rm ^{22}O$   &965          &0.2940$\pm$0.0171        &3.0403$\pm$0.0925  &1123$\pm$24$^a$\\
$\rm ^{23}O$   &960          &0.2557$\pm$0.0197        &3.2900$\pm$0.1347  &1216$\pm$41$^a$  \\
$\rm ^{24}O$   &965          &0.2193$\pm$0.0174        &3.5813$\pm$0.1511  &1318$\pm$52      \\
\\
\hline
\\
$\rm ^{18}F$   &975          &0.2412$\pm$0.0300        &3.0917$\pm$0.2123  &1100$\pm$50   \\
$\rm ^{19}F$   &985          &0.3476$\pm$0.0350        &2.6391$\pm$0.1439  &1043$\pm$24   \\
$\rm ^{20}F$   &950          &0.2686$\pm$0.0073        &3.0605$\pm$0.0425  &1113$\pm$11   \\
$\rm ^{21}F$   &1000         &0.3056$\pm$0.0110        &2.9142$\pm$0.0539  &1099$\pm$12       \\
$\rm ^{23}F$   &1020         &0.2928$\pm$0.0075        &3.0482$\pm$0.0398  &1148$\pm$16\\
$\rm ^{24}F$   &1005         &0.2385$\pm$0.0083        &3.4085$\pm$0.0609  &1253$\pm$23 \\
$\rm ^{25}F$   &1010         &0.2304$\pm$0.0075        &3.4954$\pm$0.0583  &1298$\pm$31   \\
$\rm ^{26}F$   &950          &0.2174$\pm$0.0149        &3.6233$\pm$0.1309  &1353$\pm$54   \\

\hline
\end{tabular}
\end{center}
\end{table}

\begin{table}
\caption{$\rm \sigma_{CC}^{p}$ is the contribution to the charge-changing cross section due to projectile protons with a proton target at energy E, using the values of oscillator constant, $\rm \alpha_p^2$, obtained in Table \ref{tab1}. The correction factor,
	$\rm \epsilon \left (=\sigma_{CC}^{exp}/\sigma_{CC}^{p} \right )$, is the ratio of the experimental value of charge-changing cross section ($\rm \sigma_{CC}^{exp}$) and $\rm \sigma_{CC}^{p}$. The last column gives the average value of $\rm \epsilon$ for the considered isotopes.}
\label{tab5}
\begin{tabular}{cccccc}
 
\\
\hline
\\
$\rm Projectile$  &$\rm E/A$      &$\rm \sigma_{CC}^{p}$  &$\rm  \sigma_{CC}^{exp}$ &$\rm \epsilon \left (=\frac{\sigma_{CC}^{exp}}{\sigma_{CC}^{p}} \right )$                            & $\rm  \epsilon_{avg}$\\
                 &$\rm (MeV)$    &$\rm (mb)$               &$\rm (mb)$               &                            &\\
               &               &$\rm (This~work)$        &                &    &   \\        
\\
\hline
\\
$\rm ^{12}C$   &926            &172.06$\pm$0.06          &214$\pm$7\cite{33}                 &1.244$\pm$0.040                      &1.297$\pm$\\     
$\rm ^{13}C$   &815            &168.78$\pm$0.07          &227$\pm$7\cite{33}                 &1.345$\pm$0.041                      &0.044\\
$\rm ^{14}C$   &889            &170.51$\pm$0.21          &222$\pm$9\cite{33}                 &1.304$\pm$0.051                      &\\
\\
\hline
\\
$\rm ^{14}N$   &924            &195.27$\pm$0.23          &280$\pm$5\cite{19}                 &1.434$\pm$0.024                      &1.455$\pm$\\     
$\rm ^{15}N$   &762            &187.59$\pm$0.22          &277$\pm$30\cite{19}                 &1.477$\pm$0.158                      &0.091\\
\\
\hline
\\
\end{tabular}
\end{table}


\begin{figure}
	\begin{center}
		\includegraphics[scale=1.05]{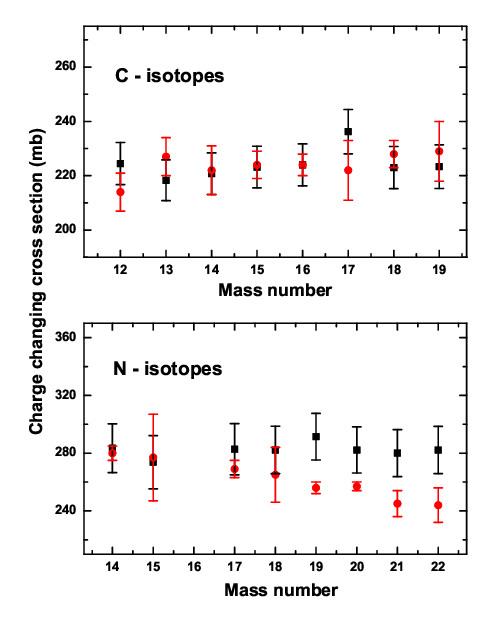}
		\caption{Charge changing cross sections  for C- and N- isotopes on a proton as a target.  The filled squares  show our theoretical predictions with the proton radii (Tables \ref{tab2} and \ref{tab3}) obtained  using the $^{12}\rm C$ target. The experimental data (filled circles) are taken from Refs.  \cite{33} and \cite{19} for C- and N- isotopes, respectively.}
		\label{fig4}
	\end{center}
\end{figure}


\subsection{Nucleus-proton charge-changing and reaction cross sections}

In order to test how far the information about the proton and neutron radii (densities), obtained in this work, suits in other situations, we now 
consider the parameter-free calculations of (i) charge-changing cross sections for $^{12-19}$\rm C \cite{33} and $^{14,15,17-22}$\rm N \cite{19} isotopes, and (ii) reaction cross sections for $^{12-19}$\rm C isotopes on a proton target in the energy range 
650-1000 MeV/nucleon. The parameters of the $NN$ amplitude are obtained in a similar way as discussed earlier. To obtain the value of the correction parameter $\epsilon$, needed in connection with the study of CCCS, we have followed the similar procedure as already discussed in Sec.\ref{sec3}A.
The required average value of the correction parameter $\epsilon$ is given 
in the last column of Table \ref{tab5}. Fig. \ref{fig4} displays the results for CCCS for $^{12-19}$\rm C and $^{14,15,17-22}$\rm N isotopes on a proton target using the 
average value of the correction parameter (Table \ref{tab5}). A fairly good agreement is seen in the case of carbon isotopes. However, the situation of nitrogen isotopes is slightly disturbing. In this case, though the results for isotopes $^{14,15,17,18}$\rm N agree with the experiment fairly well, the isotopes $^{19-22}$\rm N show large deviations between theory and experiment. In terms of the radii (densities), we find that the proton radii (densities) of $^{12-19}$\rm C, as obtained using the $^{12}$\rm C target, suit also in the case of a proton as a target. Unfortunately, this is not the case with $^{19-22}\rm N$ isotopes. It seems that one requires smaller values of the projectile proton radii in order to account for the $^{19-22}$\rm N data on a proton target. At present, there seems no valid reason in support of the expected deviations in the deduced proton radii (densities) for neutron rich nitrogen isotopes, $^{19-22}$\rm N. For if we take indications from the works of Aumann \textit{et al.} \cite{34,35} on knock-out reactions that a proton as a target can explore both the inner and surface regions of the projectile, whereas a complex target is more sensitive to the surface, we could only infer that the shape of the projectile proton density distribution, which is not the subject of the present study, may play a role in understanding the source of discrepancy in the case of $^{19-22}$\rm N isotopes; It might happen that the proton density distribution in C-isotopes, having even number of protons, is better accommodated  in the use of SDHO density as compared to that in N-isotopes having odd number of protons. A similar discussion is given in Ref.\cite{19}.

\begin{figure}
	\begin{center}
		\includegraphics[scale=1.00]{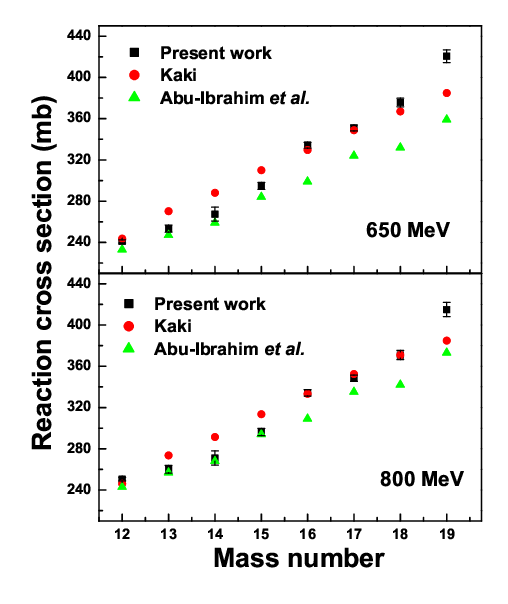}
		\caption{The predicted reaction cross sections for C-isotopes on a proton target at 650 and 800 MeV/nucleon, using our deduced proton (Tables \ref{tab2} and \ref{tab3}) and neutron (Table \ref{tab4}) radii.  The values denoted by Kaki  and Abu-Ibrahim \textit{et al.} are taken from Refs. \cite{37} and \cite{36}, respectively.}
		\label{fig5}
	\end{center}
\end{figure}

Unfortunately, no experimental data are available for the reaction cross sections ($\sigma_{R}$) of exotic neutron rich isotopes on a proton target at 
any incident energy, we have only predicted the $\sigma_{R}$ for $^{12-19}$\rm C-proton reaction at some specific energies, say 650 and 800 MeV/nucleon.  
The results of these calculations are presented in Fig. \ref{fig5}. This figure also contains the results of other calculations \cite{36,37} for comparison. Due to lack of experimental data, though no comments could be made on overall merits/demerits of the results presented in Fig. \ref{fig5}, the only result that needs attention in the present calculations is the one- neutron halo structure of $^{19}\rm C$, reflected through large increase in its $\sigma_{R}$ value not present in the other results. However, it should be mentioned that the assessment of the suitability of our extracted matter radii (densities) is a matter of future experiments on charge-changing and reaction cross sections involving different targets at different energies. 


\section{Summary and Conclusions}
\label{sec4}

In this work we have presented a theoretical study of the charge-changing and interaction cross sections for beryllium, boron, carbon, nitrogen, oxygen,
and fluorine isotopes on $^{12}$\rm C at medium energies (740-1050 MeV/nucleon), using the correlation expansion for the Glauber model S-matrix.
The densities of the colliding nuclei are obtained using the Slater determinants consisting of the harmonic oscillator single-particle wave 
functions, and the basic (input) of the Glauber model, the $NN$ amplitude, takes care of the nuclear in-medium effects.
Our main concern in this work is to extract the proton, neutron, and matter radii of the nuclei under consideration. In the first step, we have outlined the procedure for calculating the CCCS and predicted the proton radii of those exotic isotopes whose experimental charge radii are not available. From the trend of our deduced proton radii, we observe a dip at N = 14 for both the nitrogen and oxygen isotopes, reflecting this new subshell closure. Next, keeping the similar values of the proton radii as obtained from the analysis of CCCS, we have obtained the neutron radii that reproduce the corresponding interaction cross sections. It is found that our deduced matter radii agree fairly 
well with the results of other studies available in the literature. The predicted proton and neutron radii have also been used to calculate the neutron skin thickness, $(r_{n}^{pt}-r_{p}^{pt})$, for all the isotopes of a given element and studied its behaviour with respect to the experimental values of one-/two-neutron separation energies. Our results indicate that the nuclei $^{11}$\rm Be, $^{19}$\rm C, $^{22}$\rm N, and $^{26}$\rm F exhibit one-neutron halo-like structure, whereas the nuclei $^{14}$\rm Be and $^{17}$\rm B favour two-neutron halo-like structure. In rest of the isotopes, our data reveal a thick neutron surface for $^{16-18}$\rm C, $^{20,21}$\rm N, $^{22-24}$\rm O, and $^{24,25}$\rm F. 

To test the suitability of the proton and neutron radii, obtained using the $^{12}$\rm C as a target, we have performed parameter-free calculations in other situations. For this, we have considered the calculations of (i) charge-changing cross sections for $^{12-19}$\rm C and $^{14,15,17-22}$\rm N isotopes, and (ii) reaction cross sections for  $^{12-19}$\rm C isotopes on a proton as a target in the energy range considered in this work. The results for CCCS are found to agree fairly well with the experimental data for carbon isotopes. In the case of CCCS for nitrogen isotopes, we observe a fairly good agreement with the experimental data for isotopes $^{14,15,17,18}$\rm N, but the isotopes $^{19-22}$\rm N show large deviations between theory and experiment. This shows that the proton radii (densities) of $^{12-19}$\rm C, as obtained using the $^{12}$\rm C target, suit also in the case of a proton as a target. Unfortunately, this is not true for $^{19-22}$\rm N isotopes, and, at present, we have no conclusive argument in support of large deviations for the said neutron rich nitrogen isotopes. However, keeping in view of the knock-out reactions \cite{34,35}, which indicate that a proton as a target can probe the entire region of the projectile, whereas a complex target is more sensitive to the surface, we should say that one needs to focus on the shape of the projectile proton density distribution instead of projectile proton radius in order to resolve the anomaly in the case of neutron rich nitrogen isotopes.
      
Unfortunately, no experimental data are available for the reaction cross sections ($\sigma_{R}$) of neutron rich isotopes for any element on a proton target in the considered energy range, we have only predicted $\sigma_{R}$ for $^{12-19}$\rm C on a proton target at some specific energies, say 650 and 800 MeV/nucleon.
The results are simply compared with the results of other works. It is, however, suggested that the assessment of the suitability of our extracted matter radii requires future experiments on both the charge-changing and reaction cross sections for similar exotic nuclei involving different targets at different energies.

\section{ Acknowledgments}

A.A.U. acknowledges the Inter-University Centre for Astronomy and Astrophysics, Pune, India for support via an associateship and for hospitality. Z.H. acknowledges the UGC-BSR Research Start-Up-Grant (No.F.30-310/2016(BSR)).


\noindent   
\begin{thebibliography}{0}
\bibitem{1} I. Tanihata \textit{et al.}, Phys. Rev. Lett. {\bf 55}, 2676 (1985). 
\bibitem{2} P. G. Hansen and B. Jonson, Europhys. Lett. {\bf 4}, 409 (1987).
\bibitem{3} I. Tanihata, H. Savajols, and R. Kanungo, Prog. Part. Nucl. Phys. {\bf 68}, 215 (2013).
\bibitem{4} L. V. Chulkov \textit{et al.}, Nucl. Phys. A {\bf 674}, 330 (2000).
\bibitem{5} T. Yamaguchi \textit{et al.}, Phys. Rev. C {\bf 82}, 014609 (2010).
\bibitem{6} A. Estrade \textit{et al.}, Phys. Rev. Lett. {\bf 113}, 132501 (2014).
\bibitem{7} S. Terashima \textit{et al.}, Prog. Theor. Exp. Phys. {\bf 2014}, 101D02.
\bibitem{8} R. Kanungo \textit{et al.}, Phys. Rev. Lett. {\bf 117}, 102501 (2016).
\bibitem{9} S. Bagchi \textit{et al.}, Phys. Lett. B {\bf 790}, 251 (2019).
\bibitem{10} S. Kaur \textit{et al.}, Phys. Rev. Lett. {\bf 129}, 142502 (2022).
\bibitem{11} S. Ahmad, A. A. Usmani, and Z. A. Khan, Phys. Rev. C {\bf 96}, 064602 (2017).
\bibitem{12} A. Bhagwat and Y. K. Gambhir, Phys. Rev. C {\bf 69}, 014315 (2004).
\bibitem{13} T. Yamaguchi \textit{et al.}, Phys. Rev. Lett. {\bf 107}, 032502 (2011).
\bibitem{14} G. W. Fan and Xu Zhan, Int. J. Mod. Phys. E  {\bf 28}, 1950070 (2019). 
\bibitem{15} I. A. M. Abdul-Magead and Badawy Abu-Ibrahim, Nucl. Phys. A {\bf 1000},121804 (2020).
\bibitem{16} M. Tanaka \textit{et al.}, Phys. Rev. C {\bf 106}, 014617 (2022).
\bibitem{17} Xiu-Fang Li, De-Qing Fang, and Yu-Gang Ma, Nucl. Sci. Tech. {\bf 27}, 71 (2016).
\bibitem{18} I. Ahmad, J. Phys. G: Nucl. Phys. {\bf 6}, 947 (1980).
\bibitem{19} JiChao Zhang \textit{et al.}, arXiv: 2404.00682v1 [Nucl-ex] 31 Mar 2024.
\bibitem{20} H. Feshbach, A. Gal, and J. Hufner, Ann. Phys., NY, {\bf 66}, 20 (1971).
\bibitem{21} A. Ozawa \textit{et al.}, Nucl. Phys. A {\bf 709}, 60 (2002); {\bf 727}, 465 (2003).
\bibitem{22} S. Ahmad, A. A. Usmani, Shakeb Ahmad, and Z. A. Khan, Phys. Rev. C {\bf 95}, 054601 (2017).
\bibitem{23} C. Xiangzhou \textit{et al.}, Phys. Rev. C {\bf 58}, 572 (1998).
\bibitem{24} H. De Vries, C. W. De Jager, and C. De Vries, Atomic Data and Nuclear Data Tables, {\bf 36}, 495  (2017).
\bibitem{25} I. Angeli and K. P. Marinova, Table of experimental nuclear ground  state charge radii: An update, Atomic Data and Nuclear Data Tables, {\bf 99}, 69 (2013).
\bibitem{26} Tao Li, Yani Luo, and Ning Wang, Atomic Data and Nuclear Data Tables, {\bf 140}, 101440 (2021 ).
\bibitem{27} S. Ahmad, D. Chauhan, A. A. Usmani, and Z. A. Khan, Euro. Phys. J. A  {\bf 52}, 128 (2016).
\bibitem{28} M. Wang \textit{et al.}, Chinese Phys. C {\bf 36}, 1603 (2012).
\bibitem{29} JI Juan-Xia \textit{et al.}, Chinese Phys. C {\bf 36}, 43 (2012).
\bibitem{30} G. Sawhney, M. K. Sharma, and R. K. Gupta, J. Phys. G: Nucl. Part. Phys. {\bf 41}, 055101 (2014).
\bibitem{31} A. Ozawa, T. Suzuki, and I. Tanihata, Nucl. Phys. A {\bf 693},32 (2001).
\bibitem{32} R. Kanungo \textit{et al.}, Phys. Rev. C {\bf 84}, 061304(R) (2011).
\bibitem{33} Y. Suzuki \textit{et al.}, Phys. Rev. C {\bf 94}, 011602(R) (2016).
\bibitem{34} T. Aumann, C. A. Bertulani, and J. Ryckebusch, Phys. Rev. C {\bf 88}, 064610 (2013).
\bibitem{35} T. Aumann \textit{et al.}, Prog. Part. Nucl. Phys. {\bf 118}, 103847 (2021).
\bibitem{36} B. Abu-Ibrahim, W. Horiuchi, A. Kohama, and Y. Suzuki, Phys. Rev. C {\bf 77}, 034607 (2008).
\bibitem{37} K. Kaki, Prog. Theor. Exp. Phys. {\bf 2017}, 093D01 (2017).


  
\end{thebibliography}
\end{document}